\documentclass[aps,pra,twocolumn,showpacs]{revtex4}
\usepackage{color, graphicx}
\usepackage{amsmath, amssymb}
\usepackage{ulem,hyperref}

\definecolor{rosa}{cmyk}{0,1,0.50,0}
\definecolor{verde}{rgb}{0,0.6,0.2}
\definecolor{naranja}{rgb}{0.8,0.4,0}

\newcommand{\fig}[1]{Fig.~\ref{#1}}
\newcommand{\eqn}[1]{Eq.~(\ref{#1})}

\begin{document}

\title{Continuous sampling of the squeezed state nonclassicality}

\author{E. Agudelo}\email{elizabeth.ospina@uni-rostock.de}\affiliation{Arbeitsgruppe Theoretische Quantenoptik, Institut f\"ur Physik, Universit\"at Rostock, D-18051 Rostock, Germany}
\author{J. Sperling}\affiliation{Arbeitsgruppe Theoretische Quantenoptik, Institut f\"ur Physik, Universit\"at Rostock, D-18051 Rostock, Germany}
\author{W. Vogel}\affiliation{Arbeitsgruppe Theoretische Quantenoptik, Institut f\"ur Physik, Universit\"at Rostock, D-18051 Rostock, Germany}
\author{S. K\"ohnke}\affiliation{Arbeitsgruppe Experimentelle Quantenoptik, Institut f\"ur Physik, Universit\"at Rostock, D-18051 Rostock, Germany}
\author{M. Mraz}\affiliation{Arbeitsgruppe Experimentelle Quantenoptik, Institut f\"ur Physik, Universit\"at Rostock, D-18051 Rostock, Germany}
\author{B. Hage}\affiliation{Arbeitsgruppe Experimentelle Quantenoptik, Institut f\"ur Physik, Universit\"at Rostock, D-18051 Rostock, Germany}

\begin{abstract}
	We report the direct -- continuous in phase -- sampling of a regularized $P$~function, the so-called nonclassicality quasiprobability, for squeezed light.
	Through their negativities, the resulting phase-space representation uncovers the quantum character of the state.
	In contrast to discrete phase-locked measurements, our approach allows an unconditional verification of nonclassicality by getting rid of interpolation errors due to fixed phases.
	To realize the equal phase distribution of measured quadratures, a data selection is implemented with quantum random numbers created by measuring the vacuum noise.
	The continuously measured squeezed field was generated in an optical parametric amplifier.
	Suitable pattern functions for obtaining the regularized $P$~function are investigated.
	The significance of detecting negativities in our application is determined.
	The sampling of nonclassicality quasiprobabilities is shown to be a powerful and universal method to visualize quantum effects within arbitrary quantum states. 
\end{abstract}
\date{\today}
\maketitle

\section{Introduction}
\label{Sec:Introduction}
	The knowledge about the classical or quantum character of different physical systems and processes is of fundamental importance for modern physics.
	Experimental preparation, manipulation, and identification of quantum effects of light are key subjects of quantum optics.
	These topics culminated in the Nobel prize 2012~\cite{Haroche,Wineland}. 
	Nowadays researchers deal with different notions of the quantumness of physical states, looking for particular signatures that clearly reveal the true quantum nature of the underlying system.
	In quantum optics an established notion of nonclassicality, introduced in  1963, is based on the Glauber-Sudarshan $P$~representation~\cite{glauber63, sudarshan63,titulaer65}.
	The impossibility of describing optical field correlations of a given state of light in terms of classical electrodynamics ruled this definition.

	General quantum correlations of radiation fields can be characterized by a full hierarchy of inequalities for observable correlation functions, which are based on the space-time dependent $P$~functional~\cite{vogel08}.
	The quantum statistics of nonclassical light, which violates classical probability theory, can be represented by a $P$~functional, that is a quasiprobability distribution showing some negativities.
	For simultaneous measurements the $P$~functional simplifies to the $n$-partite $P$~function of the global quantum state,
	\begin{equation}
	\hat\rho =\int d^{2n}\boldsymbol\alpha\, P(\boldsymbol\alpha)\,|\boldsymbol\alpha\rangle\langle \boldsymbol\alpha|,
	\end{equation}
	written as a pseudo-statistical mixture of coherent states~\cite{glauber63, sudarshan63}, where the vectors $|\boldsymbol\alpha\rangle=|\alpha_1,\dots,\alpha_n\rangle$ denote multi-mode pro\-ducts of coherent states.
	
	Traditionally, this definition is not experimentally applicable due to the strong singularities occurring in the $P$~function for many physical systems including some single-mode states, e.g., squeezed light.
	As a consequence other phase-space representations had gained an increasing recognition.
	Due to its regularity and operational relevance in quantum state tomography~\cite{smithey93, smithey93-2} the favorite quasiprobability is the Wigner function.
	It is experimentally accessible and broadly calculated for different systems, e.g., quantum light, molecules and trapped atoms~\cite{dunn95, leibfried96, ourjoumtsev06, deleglise08,MKNPE11}.
	Despite its benign behavior, this function is less sensitive than the $P$~function to probe quantum effects.
	
	During the study of quantum correlations a particular interest was developed in working within the continuous-variable regime~\cite{korolkova, horodecki, walborn, bowen, josse}.
	Quantum information processing and communication are also developed on a continuous-variable base~\cite{LB99, braunstein05}, since for the quadrature amplitudes, which are the analogues of position and momentum, well-known modulation and detection techniques exist.
	Along with that, squeezed states are an essential tool of research in quantum physics.
	Due to the reduced noise they are useful in optical metrology~\cite{caves} and optical communications~\cite{yamamoto}.
	They have a well-behaved and positive Wigner function, but a highly irregular $P$~function.
	The general dichotomy of different established quasiprobabilities~\cite{cahill69} is that they are less sensitive for certifying quantum effects if they become more regular.  
	
	Nowadays, these conflicting properties, regularity and sensitivity to nonclassicality, can be jointly realized using nonclassicality quasiprobabilities~\cite{kiesel10}.
	These distributions visualize nonclassical states through their negativities.
	Experimental reconstructions of nonclassicality quasiprobabilities have been performed for a squeezed state measured for a discrete set of phases~\cite{kiesel11a}, requiring additional interpolations (for details see the supplement to~\cite{kiesel11a}).
	For the single-photon-added thermal state~\cite{kiesel11} on the other hand, phase properties need  not be considered in view of the radial symmetry of this state.
	The regularization method was generalized for multi-mode radiation fields~\cite{agudelo13} to uncover any simultaneous quantum correlation among the modes in the sense of the Glauber-Sudarshan notion.
	In some cases, the knowledge of the complete regularized $P$~function is not even required.
	If a regularized marginal of the $P$~function shows negativities, then the nonclassicality of the underlying quantum state is certified~\cite{kiesel12a,A93}. 
	
	In the present paper the demonstration of nonclassicality through regularized quasiprobabilities is carried out for the first time for experimentally generated states with continuous rather than discrete phase measuring.
	The limitation of the phase-locked detection with a finite number of phases is overcome by our random phase approach.
	This experimental technique allows us to record data for every relative phase between signal and local oscillator (LO). 
	Quantum random numbers are generated in order to select a uniformly distributed collection of data to sample the filtered $P$~function.
	Our continuous sampling method leads to a conclusive demonstration of the quantum character of general physical states in phase-space, as we demonstrate for the phase-sensitive squeezed vacuum state.
	The regularized $P$~function clearly displays highly significant negative values, even for a rather small amount of data.
	
	The paper is structured as follows.
	We present the general regularization method of the $P$~function for any multi-mode radiation field in Sec.~\ref{Sec:RegularizedPfun}.
	A comparison of measurements with preassigned and continuous phases is performed in Sec.~\ref{Sec:EnhanceCompare}.
	In Sec.~\ref{Sec:Experiment} the experimental setup and the continuous phase measurement technique are described.
	The nonclassicality quasiprobability is analyzed in Sec.~\ref{Sec:Results} together with the direct sampling functions and the statistical analysis of the significance of negativities in dependence on the filter parameters and the number of quadrature data.
	A summary and conclusions are given in Sec.~\ref{Sec:Conclusion}.

\section{Nonclassicality from the regularized $P$~function}
\label{Sec:RegularizedPfun}
	Firstly, we briefly recall the theory of nonclassicality quasiprobabilities.
	Eventually, we introduce a family of regularizing $q$-parametrized filter functions.
	This class includes in the limiting cases the well established $s$-parametrized quasiprobabilities ($q=2$) and an analytical but non-invertible filter ($q \to\infty$) too.
	Finally, the reconstruction of regularized phase-space representations is outlined.

\subsection{Nonclassicality quasiprobabilities}
	The starting point for constructing a regular $P$~function is to calculate its associated characteristic function (CF).
	The CF and the $P$~function are Fourier conjugate phase-space distributions.
	The former can be sampled directly from experimental data obtained from balanced homodyne detection (BHD).
	The $n$-mode CF, $\Phi(\boldsymbol\beta)$ in terms of the displacement operator is defined as
	\begin{equation}
	 \Phi(\boldsymbol\beta)=\left\langle\prod_{j=0}^{n} :\hat D_j(\beta_j):\right\rangle,
	\end{equation}
	with $:\hat D_j(\beta_j):=e^{\beta_j\hat a_j^\dagger} e^{-\beta_j^\ast \hat a_j}$ being the $j$-th mode, normally ordered displacement operator.
	The CF can be written in terms of the single-mode quadrature operators, $\hat x_j(\varphi_j)=\hat a_j e^{i\varphi_j}+\hat a_j^\dagger e^{-i\varphi_j}$, as
	\begin{equation}
	 \Phi(\boldsymbol\beta)=\left\langle e^{|\boldsymbol\beta|^2/2}\prod_{j=0}^{n} e^{i|\beta_j| \hat x_j(\pi/2-\theta_j)}\right\rangle,
	\end{equation}
	where $\beta_j=|\beta_j|e^{i\theta_j}$.
	Independent of the singularities of a $P$~function its CF is always a continuous function as well as bounded by $|\Phi(\boldsymbol\beta)|\leq e^{|\boldsymbol\beta|^2/2}$.

	A proper filtering is needed in order to easily visualize the nonclassicality of a quantum state through negativities in the regularized quasiprobabilities, rather than using the complex valued CFs.
	The filtered characteristic function (FCF) is the product~\cite{kiesel10}
	\begin{equation}
		\Phi_\Omega(\boldsymbol\beta;w)=\Phi(\boldsymbol\beta) \Omega_{\boldsymbol{w}}(\boldsymbol\beta).
	\end{equation}
	Without loss of generality the vector of real parameters, so-called filter width $\boldsymbol{w}$, can be chosen as $\boldsymbol{w}=w(1,\dots,1)^{\rm T}$, cf.~\cite{agudelo13}.
	We consider the multi-mode filter function as a product of single-mode filter functions,
	\begin{equation}\label{Eq:ProdFilter}
		\Omega_{\boldsymbol{w}}(\boldsymbol\beta) = \prod_{j=0}^{n} \Omega_w(\beta_j).
	\end{equation}
	The construction of $\Omega_w$ will be considered in the next subsection~\ref{Subsec:FilterFamilies}.
	The product filter, i.e. uncorrelated, in Eq.~\eqref{Eq:ProdFilter} is a practical tool to identify any kind of nonclassicality in the multi-mode $P$~function, even quantum correlations between different modes.
	For example, it allowed us to directly recognize quantum correlations in systems when other established methods fail~\cite{agudelo13}.
	Finally, we present the nonclassicality quasiprobability $P_{\Omega}(\boldsymbol\alpha; w)$ as the inverse Fourier transform of the FCF,
	\begin{align}\label{Eq:PQC-Def}
		P_{\Omega}(\boldsymbol\alpha; w)= \frac{1}{\pi^{2n}}\int d^{2n}\boldsymbol\beta\, e^{\boldsymbol\alpha\boldsymbol\beta^*-\boldsymbol\alpha^*\boldsymbol\beta}\, \Phi_\Omega(\boldsymbol\beta;w).
	\end{align}
	
	It has been shown in~\cite{kiesel10} that a proper filter function $\Omega_{\boldsymbol{w}}(\boldsymbol\beta)$ must satisfy three conditions.
	Firstly, the FCF is required to be a rapidly decaying function, which yields that $P_\Omega$ is well behaved, cf. also~\cite{K66}.
	Secondly, the filter has to have a nonnegative Fourier transform.
	This implies that any negativity in $P_{\Omega}$ corresponds to an authentic nonclassicality in the Glauber-Sudarshan $P$~function.
	Thirdly, for any nonclassical state exist values $w$ and $\boldsymbol\alpha$ for which $P_{\Omega}(\boldsymbol\alpha; w)<0$.
	The latter condition is fulfilled if $w\to\infty$ implies $P_{\Omega}(\boldsymbol\alpha; w)\to P(\boldsymbol\alpha)$, i.e., $P_{\Omega}$ converges to the original $P$~function.
	Since the filter is chosen independently of the prepared state, this is also a universal method to uncover quantum correlations of any physical system composed of harmonic oscillators.
	
\subsection{$q$-parametrized filters}\label{Subsec:FilterFamilies}
	A definition of a filter is given in terms of autocorrelation functions of the map $\omega_{w}(\beta)$~\cite{kiesel10}:
	\begin{align}\label{Eq:AutocFun}
		\Omega_{w}(\beta) = \int d^2\beta' \omega^\ast_{w}(\beta')\omega_{w}(\beta+\beta').
	\end{align}	
	To be a proper filter, the function $\omega_w(\beta)e^{u|\beta|^2}$ has to be square integrable for any $u>0$.
	Our considered family of $\omega_{w}(\beta)$ functions is given by
	\begin{align}\label{Eq:SingleModeFilter}
		\omega_{w,q}(\beta) = \frac{\sqrt[q]{2}}{w} \sqrt{\frac{q}{2\pi\,\Gamma(2/q)}}\, e^{-(|\beta|/w)^q},
	\end{align}
	introducing the parameter $q$ with $2<q<\infty$.
	In the limiting case $q\to2$, the function is a Gaussian and an appropriate choice of $w$ will lead to the well known $s$-parametrized quasiprobabilities~\cite{cahill69-2,AW68}.
	Among other properties, it has been shown that this filter for $2<q<\infty$ suppresses the exponentially rising behavior of $e^{|\beta|^2/2}$ and it belongs to the class of invertible filters, since $\Omega_{w,q}(\beta)$ has no zeros.
	However, the autocorrelation function Eq.~\eqref{Eq:AutocFun} must be determined numerically.
	Considering the limit when $q$ tends to $\infty$, the function $\omega_{w,\infty}(\beta)$ represents two radially symmetric Heaviside functions~\cite{kiesel12}.
	The nonclassicality filter Eq.~\eqref{Eq:AutocFun} has in this case an analytical solution~\cite{kuehn}:
	\begin{align}\label{Eq:NoninvertibleFilter}
		\Omega_w^{\infty}(\beta)=\frac{2}{\pi}\left[\arccos\left(\frac{|\beta|}{2w}\right)-\frac{|\beta|}{2w}\sqrt{1-\frac{|\beta|^2}{4w^2}}\right]\Theta\left(\frac{|\beta|}{4w}\right).
	\end{align}
	The rectangular function $\Theta(|\beta|/4w)$ is defined as $1$ for $|\beta|/4w\leqslant 1/2$ and $0$ otherwise.
	
	Previously, the condition of invertibility was included in the list of requirements, establishing that the filter function should not contain any zeros to avoid losing information about the state.
	If $\Omega_w(\beta)=0$ for some $\beta$ then the corresponding regularized $P$~function will not uniquely represent the quantum state.
	Due to this truncation, some quantum states cannot be distinguished from each other.
	However, if the resulting filtered distribution shows negativities, the physical state under study is nonclassical.
	The completeness condition, $w\to\infty$ implies $P_\Omega\to P$, still holds and guarantees a vanishing information loss in this limit.
	Due to the analytical solution in Eq.~\eqref{Eq:NoninvertibleFilter}, this filter is more convenient for numerical calculations.
	Moreover, on this basis, an analytic and invertible filter was constructed recently which preserves the full information on the quantum state~\cite{kuehn}.

\subsection{Sampling of the regularized $P$~function}
	The values of the regularized $P$~function and its standard deviation can be directly estimated from the experimental data.
	As given in Eq.~\eqref{Eq:ProdFilter}, a general multi-mode filter function is a product of single-mode ones.
	Hence, the nonclassicality quasiprobability $P_\Omega(\boldsymbol\alpha;w)$ can be written through the sampling formula
	\begin{equation}\label{Eq:SampledP}
		P_\Omega(\boldsymbol\alpha;w)=\frac{1}{N} \sum_{j=1}^{N}\prod_{k=1}^{n}\,f_{\Omega}(x_k[j],\varphi_k[j],\alpha_k;w),
	\end{equation}
	where $f_{\Omega}$ denotes the pattern function~\cite{kiesel11a,agudelo13}.
	We consider an ensemble of $N$ measured quadrature values $\{(x_k[j],\varphi_k[j])_{k=1}^n\}_{j=1}^N$.
	The index $k$ denotes the mode and $j$ numbers the measured values for the quadrature $x_k[j]$ and its corresponding phase $\varphi_k[j]$.
	The direct estimation of the physical quantities and their uncertainties consists in the evaluation of the pattern functions, which, in our case, reduces to~\cite{kiesel11a}
	\begin{align}\label{Eq:patternfct}
		f_\Omega(x,\varphi,\alpha;w)=\frac{1}{\pi}\int_{-\infty}^{+\infty} &db\, |b|\, e^{b^2/2}e^{ibx}\, \Omega_w(b)\times\\
		& e^{2 i |\alpha| b \sin\left[\arg(\alpha) -\varphi-\pi/2\right]}.\nonumber
	\end{align}
	It is important to note that $f_\Omega(x,\varphi,\alpha;w)$ has to be calculated only once and can be subsequently applied to any set of measured data.
	The sampling error of the estimated $P_\Omega$ can be naturally obtained by the mean square deviation of the numbers $f_\Omega(x_k,\varphi_k,\alpha;w)$.
	It establishes an upper bound error estimate that might be refined via a more detailed analysis.

	In order to provide the statistical significance of the negativity of the sampling we estimate the statistical uncertainties.
	The significance $S$, being a function of the filter width $w$ and the number of quadrature points $N$, is defined as the optimized ratio between the estimated value of $P_\Omega$ in Eq.~\eqref{Eq:SampledP} and its standard deviation $\sigma(P_\Omega)$,
	\begin{equation}
		S(\boldsymbol\alpha;w,N) = -\frac{P_\Omega(\boldsymbol\alpha;w)}{\sigma(P_\Omega(\boldsymbol\alpha;w))}.
		\label{Eq:signif}
	\end{equation}
	A positive value, $S(\boldsymbol\alpha;w,N)>5$, implies a significant probe of quantum properties via the quasiprobability $P_{\Omega}$ at the multimode phase-space point $\boldsymbol\alpha$.

	In order to minimize the computational effort, we take into account some Fourier techniques to perform the calculations of the pattern function Eq.~\eqref{Eq:patternfct} faster.
	First let us define the parameter $\xi=x+2 |\alpha|\sin\left[\arg(\alpha)-\varphi-\pi/2\right]$ and the function
	\begin{equation}\label{Eq:Chifct}
		\chi(\xi;w)=\frac{1}{\pi}\int_{-\infty}^{+\infty} db\, |b|\, e^{b^2/2}{e}^{ib\xi}\, \Omega_w(b),
	\end{equation}
	which is identical to the pattern function with fewer parameters,
	\begin{equation}
		f_\Omega(x,\varphi,\alpha;w)\equiv \chi(\xi;w).
	\end{equation}
	Note that $w$ attains an arbitrary but fixed value and the measured quadratures only enter in the pattern function through the parameter $\xi$.
	Therefore the function $\chi(\xi;w)$ is calculated in advance to the data processing. 
	It is noteworthy that the filter needed for sampling the Wigner function would just compensate the inverse Gaussian term in the pattern function~(\ref{Eq:patternfct}), which is still divergent in this case~\cite{kuehn}.
	Ergo, the sampling method is not possible for the Wigner function.

	The fast decay of the filter for increasing $b$, see Eq.~\eqref{Eq:patternfct} together with the filter requirements below Eq.~\eqref{Eq:PQC-Def}, is considered for the sampling of the nonclassicality quasiprobability. 
	In this case the situation improves considerably and we can well approximate the pattern function by setting it zero for all $|b|>b_c$, for sufficiently large values of $b_c$.
	The filters are fixed functions with scaled argument, $\Omega_w(\beta)=\Omega_1(\beta/w)$, as a consequence $b_c$ scales with $w$.
	In our calculations we choose $b_c=8w$, so that in all the cases the magnitude of the filter becomes less than $10^{-100}$, using double-precision floating-point numbers.
	Hence, the Fourier transform of $\chi(\xi;w)$ has a bounded support and it is a band-limited function.
	The Nyquist-Shannon sampling theorem~\cite{jerri} established that
	\begin{equation}\label{Eq:AproxChi}
		\chi(\xi;w)=\sum_{m=-\infty}^{\infty}\,\chi\big(\text{\scriptsize $\frac{\pi m}{b_c}$};w\big)\,{\rm sinc}(b_c\xi-\pi m).
	\end{equation}

	On this basis, the procedure of sampling the regularized $P$~function is as follows.
	First, the filter is calculated for a broad set of parameters $b$, e.g., in the interval $[-b_c,b_c]$.
	Then the Fourier integral Eq.~\eqref{Eq:Chifct} has to be evaluated at the discrete set of points $\pi m/b_c$.
	Afterwards the measured quadratures are inserted into Eq.~\eqref{Eq:AproxChi} in the parameter $\xi$.
	Finally, one can calculate the nonclassicality quasiprobability directly as the empirical mean of the pattern function.

	This simple procedure is solely applicable if the quadrature phases are measured and assessed in the full $2\pi$ interval forming a random uniform distribution.
	The case where only a fixed number of phase angles could be measured was tackled in~\cite{kiesel11a}.
	In such a scenario, an additional phase interpolation is indispensable which leads to additional systematic errors, for details see the supplement to~\cite{kiesel11a}.
	Here, such an additional treatment becomes superfluous -- as it is discussed in the following section.

\section{Random versus phase-locked measurements}
\label{Sec:EnhanceCompare}
	Measurement systems inherently include systematic errors, i.e., errors introduced by inaccuracy and not determined by chance.
	Their estimation typically requires a prior knowledge about the state.
        For detailed studies of experimental errors in the context of reconstruction of quantum states and quasiprobabilities we refer to~\cite{kiesel11a,kiesel2008,schwemmer15}.
	Typically, an uniformly distributed phase is needed for quantum state reconstruction~\cite{vo-ri1989}. 
	If this is not the case,  misinterpretations of quantum effects may occur, as recently discussed for Bell inequalities~\cite{kupczynski}.
        
        In the following we demonstrate that continuous phase measurements (CPM) yield, with a high significance, a reconstruction of quasiprobabilities of any quantum state.  
        In Secs.~\ref{Sec:Experiment} and~\ref{Sec:Results} we will apply this method to experimental data. 
	The advantage of this technique consist in the fact that it does not require phase interpolations as in the phase-locked measurements (PLM), e.g., cf.~\cite{kiesel11a}.
	To demonstrate the systematic errors in PLM, let us consider a displaced squeezed state. 
	This state is \textit{exotic} enough to show the strength of the CPM technique.
	The considered displaced state is represented by $|\zeta;\alpha_0\rangle=\hat D(\alpha_0)\hat S(\zeta)|{\rm vac}\rangle$, where the displacement operator is $\hat D(\alpha_0)=\exp[\alpha_0\hat a-\alpha_0^\ast\hat a]$ and the squeezing operator is $\hat S(\zeta)=\exp[(\zeta\hat a^{\dagger 2}-\zeta^\ast\hat a^2)/2]$.
		\begin{figure}[ht]
		\centering
		\includegraphics*[scale=0.35]{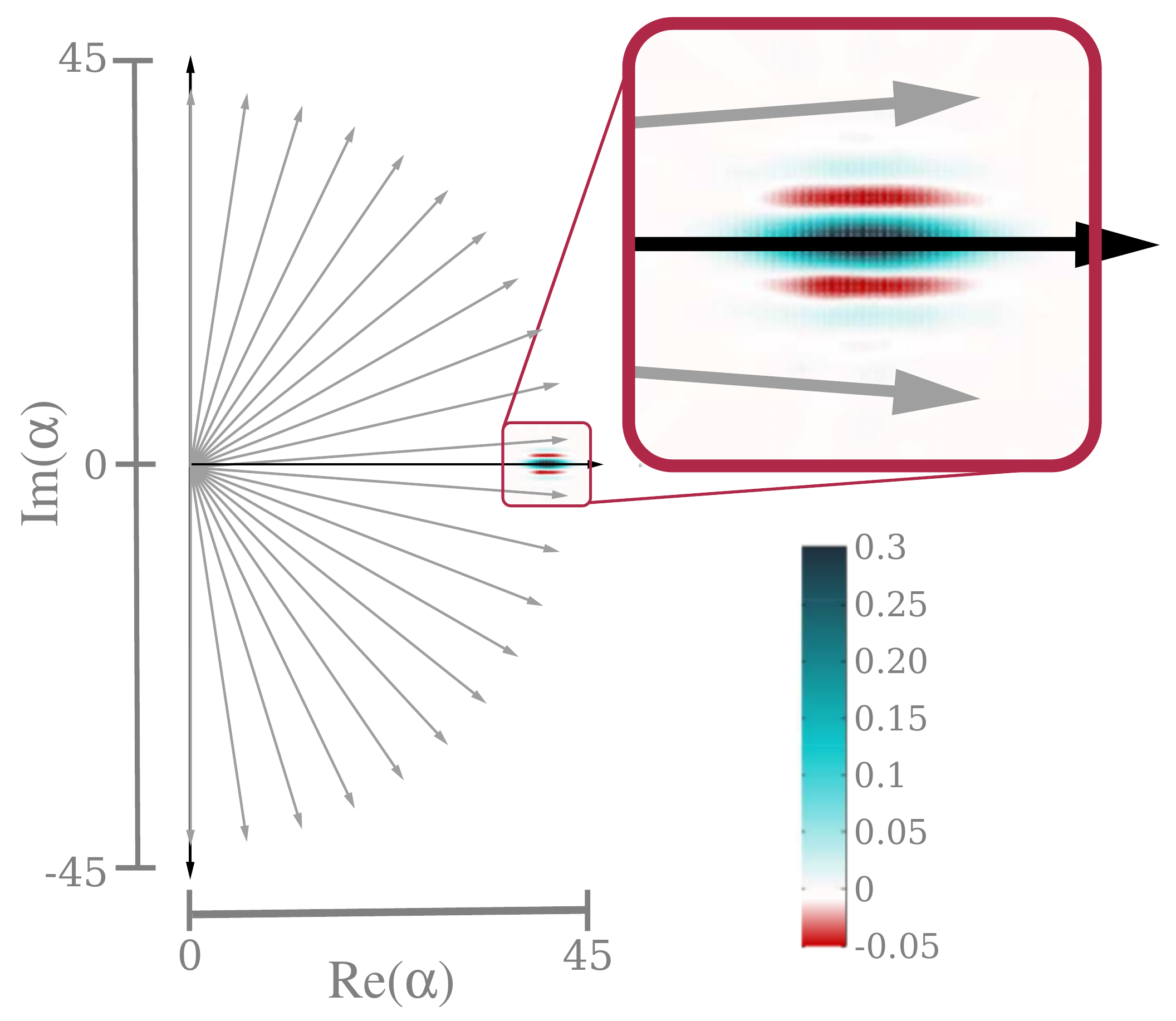}
		\caption{(Color online) 
			Density plot of the regularized $P_{\Omega}(\alpha;w)$~function for a single-mode displaced squeezed state $|\zeta;\alpha_0\rangle$.
			The squeezed state and the filter (Eqs.~\eqref{Eq:AutocFun} and ~\eqref{Eq:SingleModeFilter}) are determined by $|\zeta|=1$, $|\alpha_0|=42$ and $q=8$, $w=1.3$, respectively.
			Gray lines represent the angles of PLM for 21 equidistant phases in the interval $[0,\pi]$.
		}\label{fig:PositionPS}
		\end{figure}
		
\begin{figure}[ht]
		\centering
		\includegraphics*[scale=0.21]{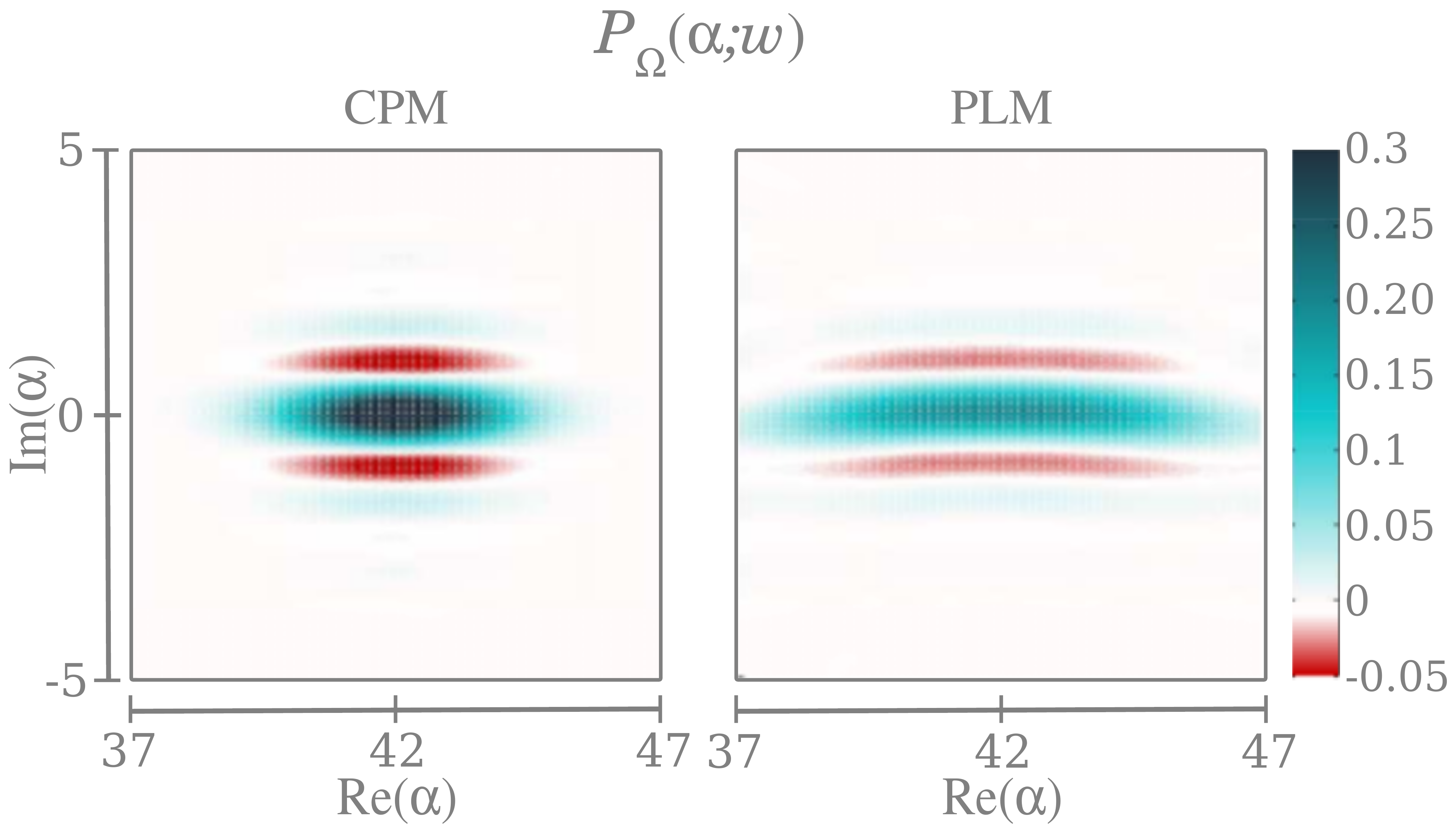}
		\caption{(Color online)
			Regularized $P_{\Omega}$~functions.
			The displaced squeezed state -- given in Fig.~\ref{fig:PositionPS} -- is reconstructed via CPM (left) and PLM (right).
		}\label{fig:RegP-CMvsPL}
		\end{figure}

	The squeezed vacuum state is horizontally displaced in phase space with an amplitude $|\alpha_0|=42$, see Fig.~\ref{fig:PositionPS}. 
	We choose 21 phases (in a $[0,\pi]$ interval), in which the quadratures are measured for the PLM.
	For the simulation we generate $N=2.1\times 10^6$ phase-quadrature pairs which corresponds to the experimental situation reported in~\cite{kiesel11a}.
	We used the invertible filter function, Eqs.~\eqref{Eq:AutocFun} and ~\eqref{Eq:SingleModeFilter}, with $q=8$ and $w=1.3$.
	The reconstructed quasiprobabilities for CPM and PLM are shown in Fig.~\ref{fig:RegP-CMvsPL}. 
	The reconstructed quasiprobability via PLM fails to be close to the true $P_{\Omega}$.
		
		\begin{figure}[ht]
		\centering
		\includegraphics*[scale=0.21]{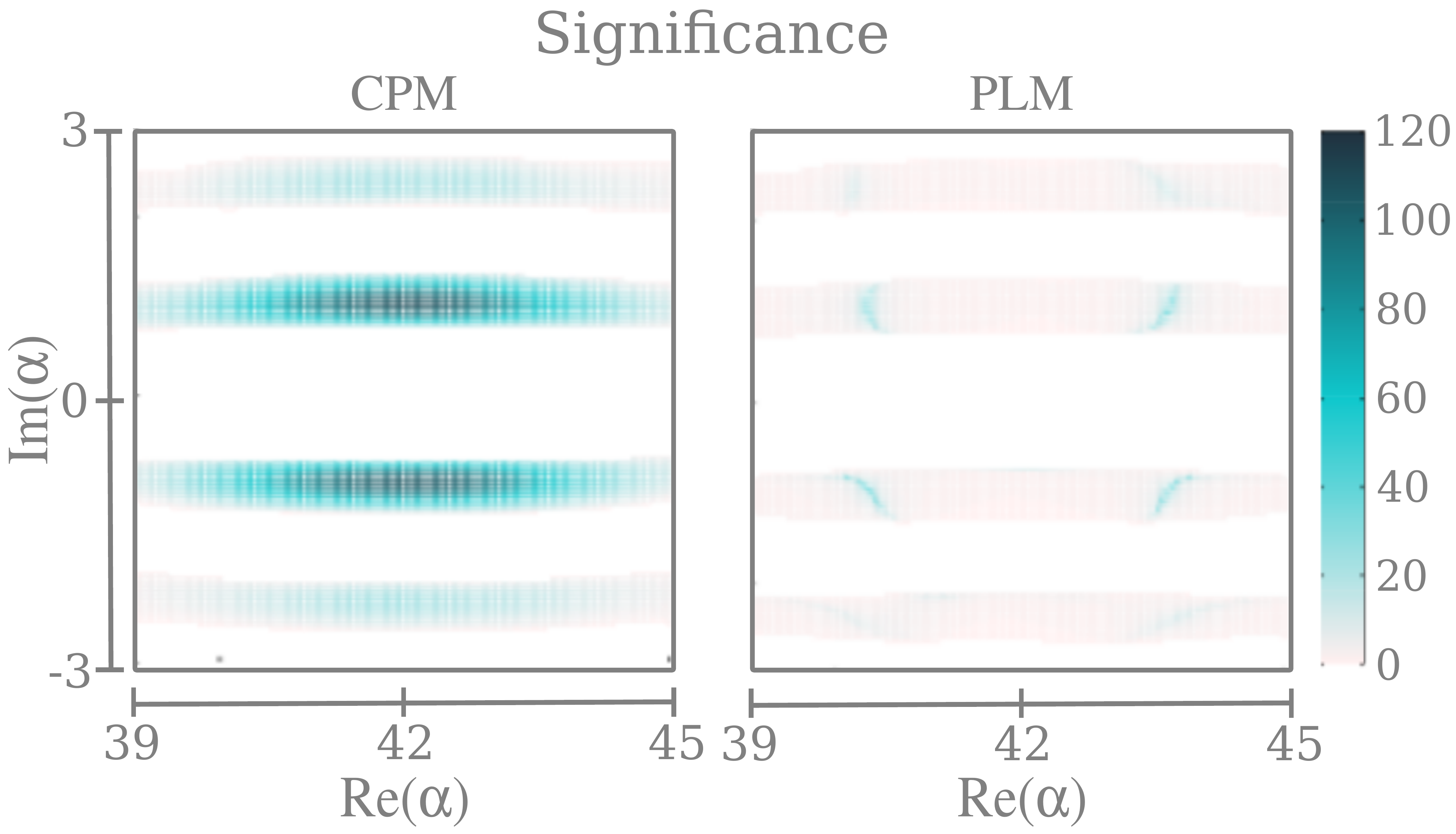}
		\caption{(Color online)
			The determined significances (Eq.~\eqref{Eq:signif}) of the negativities of the reconstructed $P_{\Omega}$~functions with a CPM (left) and a PLM (right) are display for $N=2.1\times 10^6$ simulated quadrature data.
			While the negativities for the CPM are highly significant, $S(\alpha;N,w)\gtrsim 100$, the PLM exhibits up to one order of magnitude lower significances.
			}\label{fig:Sig-CMvsPL}
		\end{figure}	
	The significances of the negativities are represented in Fig.~\ref{fig:Sig-CMvsPL}, where one can see that for CPM they are an order of magnitude bigger than for PLM.
	The latter significances of the PLM include the systematic error due to the required phase interpolation.
	Hence, the reconstruction of the regularized $P$~function through CPM is a more precise and significant method compared with the standard PLM.
	In the following, we provide a modification of BHD to continuously record the phase dependence of the quadratures.
	
\section{Experimental realization}
\label{Sec:Experiment}

\subsection{Experimental setup}
\begin{figure}[ht]
		\centering
		\includegraphics*[width=8.00cm]{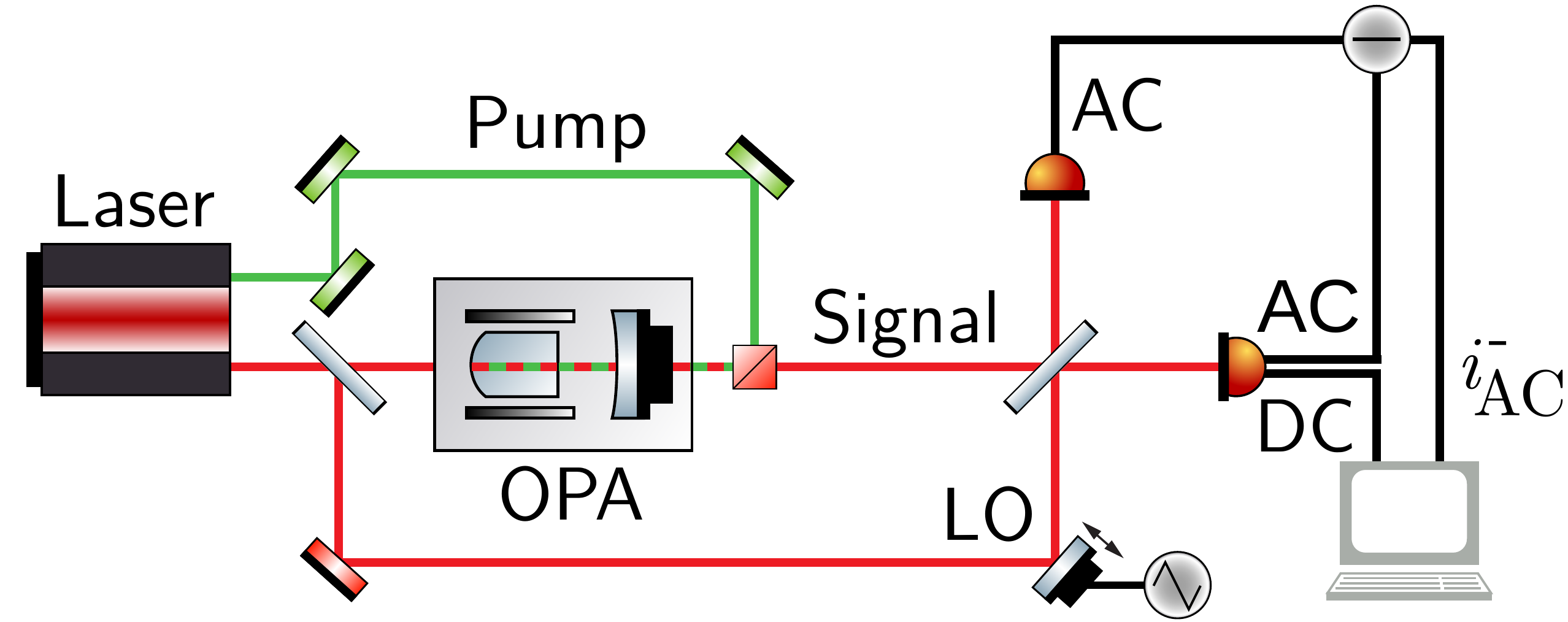}
		\caption{(Color online) Experimental setup for the generation and measurement of a squeezed state.
			The squeezed field was generated in a linear, hemilithic cavity with an 11\,\rm{mm} long, 7\% magnesium oxide-doped lithium niobate (7\%MgO:LiNbO$_{3}$) crystal, which was pumped with a 290\,\rm{mW} laser beam at 532\,\rm{nm} and which yields a strength of 3.1\,\rm{dB} squeezing at 1064\,\rm{nm}.
			A BHD (visibility 98\%, quantum efficiency 90\%, LO power 1.23\,\rm{mW}) is set up for the measurement.
			The phase of the LO was changed continuously, by applying a triangular alternating voltage onto the piezo, responsible for the mirror position.
		}\label{fig:ExpSetup}
	\end{figure}

	The squeezed field to be investigated was generated in a hemilithic, standing wave, non-linear cavity.
	This cavity was used as an optical parametric amplifier (OPA) with a $\chi^{(2)}$ nonlinear medium -- an 11\,\rm{mm} long, 7\% magnesium oxide-doped lithium niobate (7\%MgO:LiNbO$_{3}$) crystal.
	The scheme of the experimental setup is shown in Fig.~\ref{fig:ExpSetup}. 
	
	The OPA is pumped with a strong laser beam (290\,\rm{mW}) at 532\,\rm{nm}, resulting in a gain of 3.3, which yields 3.1\,\rm{dB} squeezing at 1064\,\rm{nm}.
	Due to an unstable lock of the pump phase, the squeezing ellipse fluctuated by $\pm\,3^{\circ}$ during the measurement.
	For the measurement of the generated state, a quantum state tomography (visibility 98\%, quantum efficiency 90\%) was implemented~\cite{smithey93,Breitenbach95}.
	The tomography phase, i.e. the optical phase of the signal with respect to the LO, was controlled by a mirror, which was mounted on a piezo actuator.
	So far this measurement setup represents the standard way of generating and measuring squeezed states.
	In the following, we will outline the phase variation which allows a direct sampling of the nonclassicality quasiprobability unrevealing the squeezing by a regular and nonclassical distribution.

\subsection{Continuous phase variation}
\label{subsec:phasevariation}
	The measurement technique of balanced homodyne detection (BHD) is an established and highly efficient tool in quantum optics. 
	However, the determination of the optical phase $\varphi$ of the LO with respect to the signal field is experimentally challenging. 
	Most frequently a small portion of the coherent laser beam, used as the LO, is co-propagating along the signal path.
	Hence, the LO is phase locked to the signal. 
	As a result there is a sinusoidal phase dependence of the mean (DC) difference current at the output of the BHD, which reads as
	\begin{align}
		i_\mathrm{DC}^-(\varphi)\propto\alpha_\mathrm{LO}\alpha_\mathrm{s}\cos\varphi
	\end{align}
	for a coherent signal field, $|\alpha_\mathrm{s}\rangle$.
	This signal provides precise information at the zeros, $\varphi=(2k+1)\pi/2$ for $k\in\mathbb Z$. 
	However, everywhere else it depends on the actual amplitude ($\alpha_\mathrm{LO},\alpha_\mathrm{s}$) of the signal and the LO and it does not depend on the phase near its extremes. 
	Usually, modulation techniques are used to mitigate this problem.
	Still, the readout accuracy is limited. 
	In such a scenario, the whole range of $0\leq\varphi<2\pi$ is sliced into a finite set of phases in a phase control loop. 
	This ensures that the relative phase between the LO and the signal for the currently measured quadrature is fixed within the accuracy interval of the phase readout.
	
\begin{figure}[ht]
		\centering
		\includegraphics*[viewport=0 250 800 580,scale=0.33]{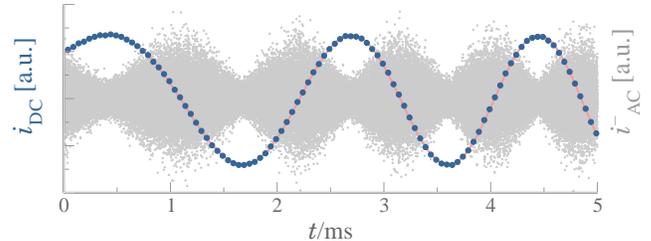}
 		\caption{(Color online)
		The measured currents.
		Blue points: Measured values of $i_{\rm DC}(t)$ while the interference phase is driven monotonically. 
		Rose line: Fitted curve according to \eqn{Eq:polyfit}. 
		Grey dot: synchronously sampled quadrature data $i^-_{\rm AC}(t)$. 
		\label{fig:PhaseRec}}
	\end{figure}

	Here we used a different approach: scanning the phase continuously by means of a movable mirror (see \fig{fig:ExpSetup}). 
	The mechanical inertia of the piezo-mirror ensures that the phase varies smoothly in time. 
	The piezo actuator was driven by a  triangular alternating voltage. 
	However, even on one slope of the triangle a constant speed cannot be guaranteed due to the nonlinear response of the piezo actuator.    
	For the subsequent estimation of the relative phase we recorded the DC current of one BHD-detector on each slope, whereat one slope is one direction of motion of the triangular drive.
	The output signal for one slope is a sine $i_{\rm DC}(t)=a\sin(\varphi(t))+b$ with a certain amplitude $a$ and an offset $b$. 
	In order to account for the smooth but nonlinear character of the phase over time we used a fourth order polynomial $\varphi(t)=gt^4+ft^3+et^2+dt+c$ and fitted the coefficients $(a,\ldots,g)$ in the expression 
	\begin{align}
		i_{\rm DC}(t)=a\sin(gt^4+ft^3+et^2+dt+c)+b,
		\label{Eq:polyfit}
	\end{align}
	to the measured signal (see. \fig{fig:PhaseRec}). 
	The resulting quality of the fit represented, e.g., by the \textit{coefficient of determination} of $R^2>99.9\%$ \footnote{The coefficient of determination is defined as $R^2=1-V_\mathrm{f}/V_\mathrm{t}$, with $V_\mathrm{t}$ being  the variance of the measured data and $V_\mathrm{f}$ being the variance of the residuals, i.e. the deviation of the measured data from the fitted curve.} indicates that the phase evolution can be described well by a fourth order polynomial in time.
	
	For technical reasons, mainly the dynamic range, the quadrature signal $i^-_{\rm AC}(t)$ has to be treated separately from the mean intensity signal $i_{\rm DC}(t)$. 
	In order to ensure that the phase information is obtained the way described above, both signals have to be sampled synchronously which actually yields the detection phase of the quadrature signal.
	In our data acquisition system we derived the sampling clocks for both signals from the same master clock.
	The quadrature data were sampled at 1~GS/s (gigasamples) whereas the DC signal was recorded at 100~kS/s.
	Subsequently the quadrature data were resampled to 20~MS/s leaving 200 data points for each sample of the DC signal.
	Moreover they were digitally band-pass filtered from 4\,\rm{MHz} to 8\,\rm{MHz}, in order to avoid technical noise.
	With this approach of varying the phase continuously an additional interpolation between discrete, locked phases becomes superfluous.

\subsection{Data processing}
	The experimental data consist in the set of  $N=2.5\times10^6$ quadrature data measured in a couple of seconds, this original set of measurements go through a post-selection process.
	First, owing to the unknown response of the piezo at the turning points of the mirror, any phase estimation at those points is not feasible.
	Those data are simply withdrawn for any subsequent estimation.
	
	In addition, due to the arbitrary choice of the start and ending time of recording the measurements, the experimental data is not uniformly distributed on phases which is a requirement of the theoretical method proposed here.
	Therefore, we proceed to statistically select a uniformly distributed phase sample based on the generation of quantum random numbers following~\cite{Gabriel} and \cite{Symul}.
	The quantum random numbers were generated from measurements of the vacuum field, which is enclosed in the radio-frequency sidebands of our local oscillator.
	Blocking the signal beam (Fig.~\ref{fig:ExpSetup}) is sufficient for measuring the quadratures of a mode in the lowest energy vacuum state.
	Quantum random numbers between zero and $2\pi$ are produced and related to the phases in the original data set (described in section \ref{subsec:phasevariation}).
	The data pairs $\{(x_j,\varphi_j)\}_{j=1}^N$ with the coincident phases are selected as the sample for the regularized $P$~function reconstruction. 

\begin{figure}[ht]
		\centering
		\includegraphics*[viewport=0 20 750 580,scale=0.33]{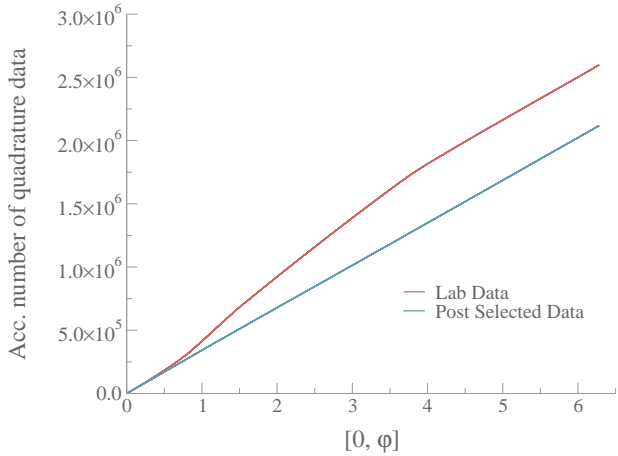}
		\caption{(Color online)
			The accumulated number of quadrature data points in the interval $[0,\varphi]$ for the experimental data set (rose) and the post-selected data set (blue).
			Statistical data selection is based on coincidences with unique quantum random numbers produced directly from vacuum states.
			The linear behavior of processed data verifies a uniform phase distribution of quadrature data points.
		}\label{fig:PhaseDist}
	\end{figure}

	The accumulated number of quadrature data depending on the phase is shown in Fig.~\ref{fig:PhaseDist}.
	We observe for the post-selected set a linearly increasing slope with increasing phase.
	This directly verifies a uniformly distributed phase, since for the underlying cumulative probability distribution holds
	\begin{align}
		{\cal P}_{\rm uniform}([0,\varphi])=\int_{0}^\varphi d\varphi'\,\frac{1}{2\pi}=\frac{\varphi}{2\pi}.
	\end{align}
	The distributions of phases in the original data set does not fulfill this uniform behaviour.
	
\section{Direct verification of nonclassicality}
\label{Sec:Results}
	In the following we reconstruct the nonclassicality quasiprobabilities of quantum states measured by BHD in terms of the pattern functions in Eq.~\eqref{Eq:patternfct}.
	These functions allow a direct sampling -- continuous in phase -- for the estimation of the quasiprobability.
	In order to probe the quality of any negativity within a sampled quasiprobability distribution, its statistical significance is of particular interest.
	We will perform such an analysis by applying our theory to the squeezed-state experiment.

\begin{figure}[ht]
		\centering
		\includegraphics*[scale=0.48]{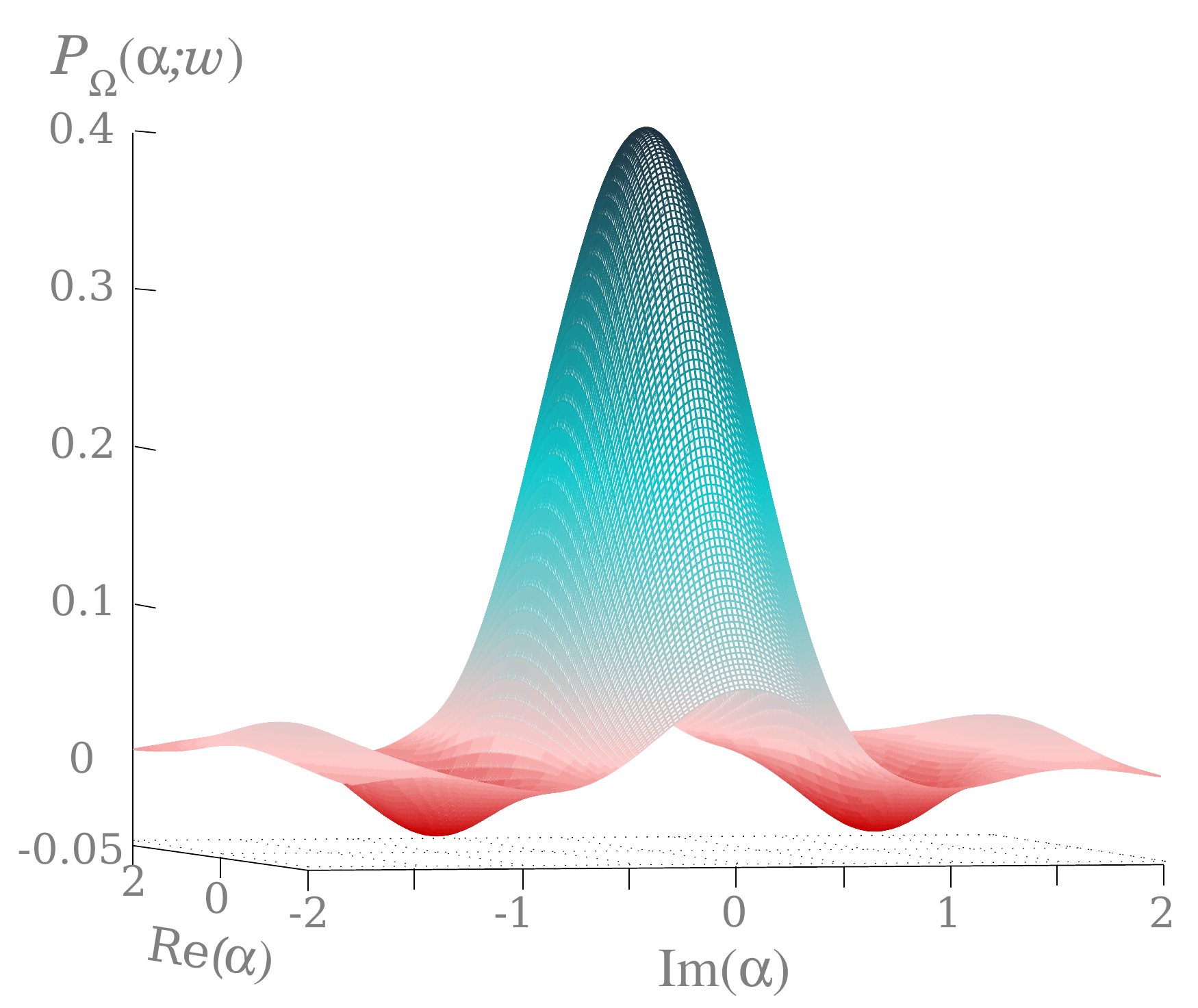}
		\caption{(Color online)
		The nonclassicality quasiprobability for the optimal set of parameters $q=\infty$ and $w=1.3$.
		In contrast to the Wigner function, this phase-space representation shows the nonclassicality of the squeezed state in terms of negativities,
		and differently to the Glauber-Sudarshan $P$~function it is a regular distribution.
		}\label{fig:3DregP}
	\end{figure}

	Before we will discuss the dependence of the filtered quasiprobability on the filter width $w$, the parameter $q$, and the number of data points $N$, we present one reconstructed phase-space representation via pattern functions from the measured $N=2\times10^6$ data points.
	The sampled regularized $P$~function in Fig.~\ref{fig:3DregP} clearly shows negativities.
	These negativities are a direct proof of the quantum nature of the state.
	As we will see later on, $q=\infty$ and $w=1.3$ represent the optimal choice of parameters in terms of significance.
		
\begin{figure}[ht]
		\centering
		\includegraphics*[viewport=0 20 750 580,scale=0.33]{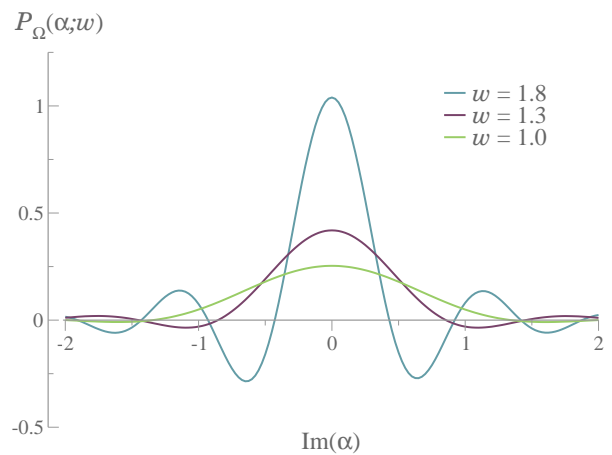}
		\caption{(Color online)
			Cut [${\rm Re}(\alpha)\equiv0$] through the nonclassicality quasiprobability $P_{\Omega}(\alpha;w)$ of the squeezed vacuum for filter widths $w=1.0,1.3,1.8$.
			These regularized $P$~functions were estimated with a $q$-parametrized filter for $q = 8$.
			Clear negativities visualize the nonclassical nature of the state.
		}\label{fig:regPs}
	\end{figure}

	Cuts of the regularized $P$~function along the axis $\rm{Im}(\alpha)$ in Fig.~\ref{fig:regPs} show the negativities for different widths $w$.
	The larger the value of $w$, the more visible the nonclassical effects are.
	Here an invertible filter ($q<\infty$) is considered and for all the points the standard deviation is smaller than the width of the line in the plot.

	In order to provide the highest statistical significance of the sampled negativities, we define the maximal significance $\Sigma$ for any point in phase space:
	\begin{equation}
		\Sigma(w,N) =\max_\alpha{S(\alpha;w,N)} =-\min_\alpha\left[\frac{P_\Omega(\alpha;w)}{\sigma(P_\Omega(\alpha;w))}\right],
		\label{Eq:signifPRIME}
	\end{equation}
	cf. Eq.~\eqref{Eq:signif}.
	For $\Sigma<0$, this metric yields the most significant nonclassical contribution of the quasiprobability.
	Hence, $\Sigma$ quantifies the exclusion of any description of the measured state in terms of classical probability theory.

\begin{figure}[ht]
		\centering
		\includegraphics*[viewport=0 20 750 580,scale=0.33]{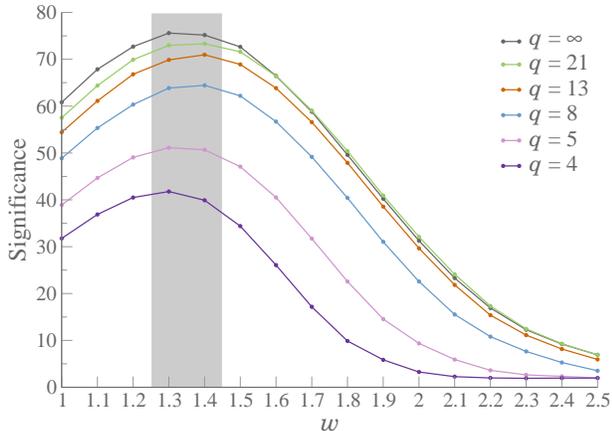}
		\caption{(Color online)
			Significance of the negativities of the sampled filtered $P$~function. 
			Different curves represent different $q$-parametrized filters -- from bottom to top: $q=4,5,8,13,21,\infty$, for
			$N=2\times10^6$.
			The region of optimal widths $w$ is highlighted in gray.
			}\label{fig:Signif-w}
	\end{figure}

	Figure ~\ref{fig:Signif-w} shows the dependence of $\Sigma(w,N)$ on the filter width $w$ for $N=2\times10^6$.
	An increasing $q$ -- describing the decay behavior of the filter $\Omega_w(\beta)$ -- results in larger statistical significance of the negativity.
	The negativities in the estimated nonclassicality quasiprobability function have significances up to 75 standard deviations. 
	Therefore it is representing an authentic demonstration of the nonclassicality present in the Glauber-Sudarshan $P$~function.

\begin{figure}[ht]
		\centering
		\includegraphics*[viewport=0 20 750 550,scale=0.33]{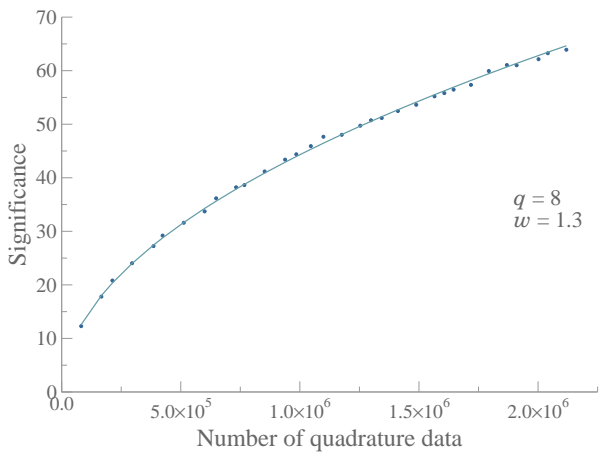}
		\caption{(Color online)
			Significance of the negativities of the sampled filtered $P$~function for different amount of quadrature data $N$.
			The autocorrelation filter, Eqs.~\eqref{Eq:SingleModeFilter} and~\eqref{Eq:AutocFun}, has the parameters $q =8$ and $w =1.3$.
			The dotted curve represents the sampled results, the full line shows the theoretically expected $\sqrt{N}$ behavior.
		}\label{fig:Signif-nd}
		\end{figure}

	The dependence of $\Sigma(w,N)$ on the sample size $N$ for fixed $w=1.3$ is investigated in Fig.~\ref{fig:Signif-nd}.
	The obtained negativities with all the available data and the filter parameter $q=8$ reach a maximum significance of 63 standard deviations. 
	We observe that even for 10\% of the available data, $N=2.5\times10^5$, significant negativities are observed. 
	In this case we have a significance of $\Sigma(1.3,2\times10^5)>20$.
	The expected reduction of sampling noise with increasing number $N$ of data points, $\Sigma(w,N)\propto\sqrt N$ for fixed $w$, is clearly visible too.

\section{Summary and conclusions}
\label{Sec:Conclusion}
	We introduced an approach of continuous phase variation for sampling regular phase-space distributions.
	The continuous scanning of the phase is an easily implementable tool, which yields highly significant results for the nonclassicality quasiprobabilities.
	Interpolation errors, being present in phase locked detection scenarios, can be overcome with our measurement strategy.
	The directly sampled phase-space quasiprobabilities lead to a verification of any nonclassicality which is present in the highly-singular Glauber-Sudarshan distribution.
	This can be done without any prior knowledge about phase-space properties of the state, which would be required for measurements at a discrete set of phases. 
	The advantages of the continuous phase measurement are demonstrated by a simulation of our method for a squeezed coherent state.

	The direct sampling of quasiprobabilities requires an equal distribution of the measured quadrature values as a function of the phase. 
	Due to nonlinearities in the electro-optical phase control, this is not directly realized in the experiment.
	To overcome this deficiency, we have implemented a data selection based on quantum random numbers, which are created by our setup.
	To prove the strength of our approach, a squeezed state with a relatively weak squeezing was realized and characterized.
	The efficiency of the method was underlined by a study of the significance of the certified quantum effects, in dependence on the number of measured data and other parameters.

	Pattern functions have been numerically constructed in advance of the data processing.
	They apply to any kind of quantum state and minimize the computational effort during the data analysis.
	The theory of the multi-mode filtering can be formulated in terms of uncorrelated filter functions.
	Thus, our single mode experiment serves as a proof-of-principle demonstration of future multi-mode schemes to probe the quantum nature of complex radiation fields.
	Therefore the presented method paves the way to visualize general quantum correlations in terms of negative values of regular, and hence experimentally accessible, phase-space distributions.

\section*{Acknowledgments}
	The authors are grateful to B. K\"uhn for enlightening discussions.
	This work was supported by the Deutsche Forschungsgemeinschaft through SFB 652 (B12 and B13).

\end{document}